\def\tr{{\rm tr}\,}
\def\Tr{{\rm Tr}\,}

\def\b{\bibitem}
\def\wt{\widetilde}
\def\be{\begin{equation}}
\def\ee{\end{equation}}
\def\bea{\begin{eqnarray}}
\def\eea{\end{eqnarray}}
\def\bml{\begin{mathletters}}
\def\eml{\end{mathletters}}
\documentstyle[aps,prb,eqsecnum,psfig,multicol]{revtex}

\newcommand{\bleq}{\ifpreprintsty
                   \else
                   \end{multicols}\vspace*{-3.5ex}{\tiny
                   \noindent\begin{tabular}[t]{c|}
                   \parbox{0.493\hsize}{~} \\ \hline \end{tabular}}
                   \fi}
\newcommand{\eleq}{\ifpreprintsty
                   \else
                   {\tiny\hspace*{\fill}\begin{tabular}[t]{|c}\hline
                    \parbox{0.49\hsize}{~} \\
                    \end{tabular}}\vspace*{-2.5ex}\begin{multicols}{2}
                    \fi}
\newcommand{\bcols}{\ifpreprintsty\else\begin{multicols}{2}\fi}
\newcommand{\ecols}{\ifpreprintsty\else\end{multicols}\fi}

\begin{document}
\draft
\def\SNG{{\em Physical Review Style and Notation Guide}}
\def\LUG {{\em \LaTeX{} User's Guide \& Reference Manual}}
\def\btt#1{{\tt$\backslash$\string#1}}%
\def\REVTeX{REV\TeX}
\def\AmS{{\protect\the\textfont2
        A\kern-.1667em\lower.5ex\hbox{M}\kern-.125emS}}
\def\AmSLaTeX{\AmS-\LaTeX}
\def\BibTeX{\rm B{\sc ib}\TeX}
 
\title{Transport properties of clean and disordered superconductors in 
matrix field theory}
\author{Lubo Zhou}
\address{Chemical Physics Program, Institute for Physical Science and 
Technology\\
University of Maryland,
College Park, MD 20742}
\author{T.R. Kirkpatrick}
\address{Institute for Physical Science and Technology, and Department 
of Physics\\
University of Maryland, 
College Park, MD 20742}

\date{\today}
\maketitle

\begin{abstract}
A comprehensive field theory is developed for superconductors 
with quenched disorder. We first 
show that the matrix field theory, used previously to
describe a disordered Fermi liquid and a disordered itinerant 
ferromagnet, also has a saddle-point solution that
describes a disordered superconductor. A general gap equation is 
obtained. We then expand about 
the saddle point to Gaussian order to explicitly obtain the 
physical correlation 
functions. The ultrasonic attenuation, number density 
susceptibility, spin density susceptibility and the electrical 
conductivity are used as examples. Results in the clean limit and 
in the disordered case are discussed respectively. This formalism is 
expected to be a powerful tool to study the 
quantum phase transitions between the normal metal state and the 
superconductor state.
\end{abstract}
\pacs{PACS numbers: }

\bcols
\section{Introduction}
\label{sec:I}

The description of disordered many-electron systems is a difficult problem 
in modern theoretical physics. 
Landau Fermi-liquid theory\cite{BaymPethick}
and then many-body perturbation theory\cite{FetterWalecka,AGD} were introduced 
to deal with this problem.
Considerable progress has been made within the framework of the latter.
On the other hand,
field-theoretic method has also been applied 
to the many-electron problem, which has certain advantages over the 
traditional technique. It is relatively easy to include the 
quenched disorder in 
the formalism with the help of replica trick. More importantly, it is 
the natural language to describe any classical or quantum phase transition. 
It allows for a straightforward 
application of the renormalization group, implementing an old program 
of describing the various phases of many-body systems in terms of stable RG 
fixed points.\cite{Shankar} So far this program has been carried out for 
clean and disordered Fermi liquids, as well as disordered 
ferromagnetic metals.\cite{Shankar_ff,us_fermions}

In the present paper we will develop a comprehensive field-theoretical method, 
or matrix field theory,\cite{us_fermions} for gapped disordered spin-singlet 
superconductors. The formalism is expected to be able to describe 
the quantum phase transition between the normal metal state and the 
superconductor state. Some 
similar techniques have been applied to describing 
spin-triplet, even-parity superconductors, but  explicit 
quantitative expressions for the Gaussian propagators could not be given
in the previous paper and thus that description cannot be said to be 
complete.\cite{Tri_SC} 
Here we will completely determine all the soft, or 
gapless correlation functions for the $S = 0$, spin-singlet case and obtain
the corresponding transport properties. 
The method can be generalized to evaluate other physical systems, 
like spin-triplet superconductors. Our results for the spin-singlet case 
coincide with earlier ones obtained by conventional methods. 
Our field theoretic methods, however, have the advantage 
that they can be easily generalized to describe quantum phase transitions. 
For example, in future publications we will use these results 
to (1) describe a metal--superconductor transition in a dirty metal, 
without integrating 
out the soft fermionic degrees of freedom\cite{Second} and 
(2) consider the same superconductor--metal transition from the 
superconducting side of the 
transition.\cite{UNP}  
This second problem is nontrivial
and interesting due to the numerous dangerous irrelevant 
variables at this phase transition.

The paper is organized as follows. In Sec.\ \ref{sec:II} we 
give a field-theoretic formulation of
the problem.
In Sec.\ \ref{sec:III} we develop the theory to construct a 
saddle-point solution, 
obtain the gap equation of superconductivity and expand to the 
second or Gaussian order about the saddle point. 
In Sec.\ \ref{sec:IV} we show how to calculate the ultrasonic 
attenuation coefficient, the number and spin density susceptibilities, 
and the electrical conductivity for both
clean and disordered
superconductors.
In Sec.\ \ref{sec:V}
we conclude with a general discussion of our results.
In Appendix we give some technical points that are used in the paper.


\section{Matrix field theory}
\label{sec:II}

\subsection{Grassmannian field theory}
\label{subsec:II.A}

In general, a system of interacting, disordered electrons will be considered. 
The partition function of the system is\cite{NegeleOrland} 
\begin{equation}
Z=\int D[\bar{\psi},\psi]\ e^{S[\bar{\psi},\psi]}\quad. 
\label{eq:2.1}
\end{equation}
Here the $\bar\psi$ and $\psi$ are Grassmann valued fields, 
and $S$ is the action including three parts:
\begin{mathletters}
\label{eqs:2.2}
\begin{equation}
S = S_0 + S_{\rm int} + S_{\rm dis}\quad, 
\label{eq:2.2a}
\end{equation}
Where, with a $(d+1)$-vector notation $x=({\bf x},\tau)$ and
$\int dx=\int_V d{\bf x}\int_{0}^{\beta} d\tau$, $S_0$ describes 
free electrons with chemical potential $\mu$,
\begin{equation}
S_0 = \int dx\sum_{\sigma}\ \bar{\psi}_\sigma (x)
\,\left(- \partial_{\tau} + \frac{\nabla^2}{2m}
       + \mu\right)\,\psi_{\sigma}(x)\quad, 
\label{eq:2.2b}
\end{equation} 
$S_{\rm int}$ describes a spin--independent two--electron interaction,
\begin{eqnarray}
S_{\rm int} &=&-\frac{1}{2}\int dx_1\,dx_2\ \sum_{\sigma_1,\sigma_2}\ 
            v({\bf x}_1 - {\bf x}_2)
\nonumber\\
          &&\times\bar{\psi}_{\sigma _1}(x_1)\,\bar{\psi}_{\sigma _2}(x_2)\,
             \psi _{\sigma_2}(x_2)\,\psi _{\sigma _1}(x_1) 
\label{eq:2.2c}
\end{eqnarray}
and $S_{\rm dis}$ describes a static random potential 
$u({\bf x})$ coupling to the electronic number density,
\begin{equation}
S_{\rm dis} = -\int dx\sum_{\sigma}\ u({\bf x})\,\bar{\psi}_\sigma (x)\,
                                             \psi_{\sigma}(x)\quad. 
\label{eq:2.2d}
\end{equation}
\end{mathletters}%

For further calculation, we assume the random 
potential $u({\bf x})$ in Eq.\ (\ref{eq:2.2d}) has variance, 
\begin{mathletters}
\label{eqs:2.3}
\begin{equation}
\left\{ u({\bf x})\,u({\bf y})\right\}_{\rm dis} = 
\frac{1}{\pi N_F \tau_{\rm e}}\ 
           \delta ({\bf x}-{\bf y})\quad,
\label{eq:2.3a}
\end{equation}
and is Gaussian distributed,
\begin{equation}
\left\{\ldots\right\}_{\rm dis} = \int D[u]\ P[u]\ (\ldots)\quad,
\label{eq:2.3b}
\end{equation}
\end{mathletters}%
$N_F$ is the density of states at the
Fermi level, and $\tau_{\rm e}$ is the elastic scattering time.
The disorder is quenched, so the replica trick\cite{ReplicaTrick} is used. With
\begin{equation}
\ln Z = \lim_{m\rightarrow 0}\ (Z^m - 1)/m\quad,
\label{eq:2.4}
\end{equation}
we consider,
\begin{equation}
{\wt Z}\equiv\{Z^m\}_{\rm dis} 
      = \int\prod_{\alpha =1}^{m} D\left[\bar{\psi}^{\alpha},
                                                         \psi^{\alpha}\right]
\exp [\tilde S\,]\quad,
\label{eq:2.5}
\end{equation}
where the corresponding action $\tilde S$ equals to
\begin{equation}
\tilde{S} = \sum_{\alpha =1}^{m}\ \left(\tilde{S}_{0}^{\,\alpha} 
                 +\tilde{S}_{\rm int}^{\,\alpha} 
+ \tilde{S}_{\rm dis}^{\,\alpha}\right)
                         \quad.
\label{eq:2.6}
\end{equation}

It is also useful to get a Fourier representation with wave vectors
${\bf k}$ and fermionic Matsubara 
frequencies $\omega_n=2\pi T(n+1/2)$ by the following transformations:
\bml
\label{eqs:2.7}
\bea
\psi_{n\sigma}({\bf x})&=&\sqrt{T}\int_0^{\beta}d\tau\ e^{i\omega_n\tau}\,
   \psi_{\sigma}(x)\quad,
\nonumber\\
{\bar\psi}_{n\sigma}({\bf x})&=&\sqrt{T}\int_0^{\beta}d\tau
   \ e^{-i\omega_n\tau}\,{\bar\psi}_{\sigma}(x)\quad,
\label{eq:2.7a}
\eea
and
\bea
\psi_{n\sigma}({\bf k})&=&\frac{1}{\sqrt{V}}\int d{\bf x}\ e^{-i{\bf k}\cdot
   {\bf x}}\,\psi_{n\sigma}({\bf x})\quad,
\nonumber\\
{\bar\psi}_{n\sigma}({\bf k})&=&\frac{1}{\sqrt{V}}\int d{\bf x}\ 
   e^{i{\bf k}\cdot{\bf x}}\,{\bar\psi}_{n\sigma}({\bf x})\quad.
\label{eq:2.7b}
\eea
\eml%
The procedure used here is similar to 
the one used in Ref.\ \onlinecite{us_fermions}, and we
refer the reader to it for further details.

\subsection{Composite variables: $Q$-matrix}
\label{subsec:II.B}

Now we integrate out the Grassmann fields and rewrite the theory in terms of 
complex-number fields. As a first step, the resulting 
model can then be approximately 
solved by using saddle-point techniques. Later fluctuations 
about the saddle point will be considered. 
First we introduce a matrix of bilinear products of the fermion fields,
\begin{eqnarray}
B_{12} &=& \frac{i}{2}\,\left( \begin{array}{cccc}
          -\psi_{1\uparrow}{\bar\psi}_{2\uparrow} &
             -\psi_{1\uparrow}{\bar\psi}_{2\downarrow} &
                 -\psi_{1\uparrow}\psi_{2\downarrow} &
                      \ \ \psi_{1\uparrow}\psi_{2\uparrow}  \\
          -\psi_{1\downarrow}{\bar\psi}_{2\uparrow} &
             -\psi_{1\downarrow}{\bar\psi}_{2\downarrow} &
                 -\psi_{1\downarrow}\psi_{2\downarrow} &
                      \ \ \psi_{1\downarrow}\psi_{2\uparrow}  \\
          \ \ {\bar\psi}_{1\downarrow}{\bar\psi}_{2\uparrow} &
             \ \ {\bar\psi}_{1\downarrow}{\bar\psi}_{2\downarrow} &
                 \ \ {\bar\psi}_{1\downarrow}\psi_{2\downarrow} &
                      -{\bar\psi}_{1\downarrow}\psi_{2\uparrow} \\
          -{\bar\psi}_{1\uparrow}{\bar\psi}_{2\uparrow} &
             -{\bar\psi}_{1\uparrow}{\bar\psi}_{2\downarrow} &
                 -{\bar\psi}_{1\uparrow}\psi_{2\downarrow} &
                      \ \ {\bar\psi}_{1\uparrow}\psi_{2\uparrow} \\
                    \end{array}\right)
\nonumber\\
&\cong& Q_{12}\quad,
\label{eq:2.8}
\end{eqnarray}
where all fields are understood to be taken at position ${\bf x}$, and
$1\equiv (n_1,\alpha_1)$ with $n_1$ denoting a 
Matsubara frequency and $\alpha$ a replica index, etc. The
matrix elements of $B$ commute with one another, and are therefore
isomorphic to classical or complex number-valued fields that we denote by 
$Q$. We use the notation $a\cong b$ for ``$a$ is isomorphic to
 $b$''.
This isomorphism maps the adjoint operation on products of fermion fields,
which is denoted above by an overbar, onto the complex conjugation of the
classical fields. We use the isomorphism to
constrain $B$ to the classical field $Q$ by means of a 
functional $\delta$ function, 
and exactly rewrite the partition
function\cite{us_fermions}
\begin{eqnarray}
\wt{Z} &=& \int D[{\bar\psi},\psi]\ e^{{\tilde S}[{\bar\psi},\psi]}
      \int D[Q]\,\delta[Q-B]
\nonumber\\
  &=& \int D[{\bar\psi},\psi]\ e^{{\tilde S}[{\bar\psi},\psi]}
      \int D[Q]\,D[{\wt\Lambda}]\ e^{\Tr [{\wt\Lambda}(Q-B)]}
\nonumber\\
  &\equiv& \int D[Q]\,D[{\wt\Lambda}]\ e^{{\cal A}[Q,{\wt\Lambda}]}\quad.
\label{eq:2.9}
\end{eqnarray}
Here ${\wt\Lambda}$ is an auxiliary bosonic matrix 
field that plays the role of 
a Lagrange multiplier, and integrates out the
fermion fields.

It is useful to expand the $4\times 4$ matrix in Eq.\ (\ref{eq:2.8})
in a spin-quaternion basis,
\begin{equation}
Q_{12}({\bf x}) = \sum_{r,i=0}^{3} (\tau_r\otimes s_i)\,{^i_rQ_{12}}({\bf x})
                 \quad
\label{eq:2.10}
\end{equation}
and analogously for $\wt\Lambda$. Here 
$\tau_0 = s_0 = \openone_2$ is the
$2\times 2$ unit matrix, and $\tau_j = -s_j = -i\sigma_j$, $(j=1,2,3)$,
with $\sigma_{1,2,3}$ the Pauli matrices. In this basis, $i=0$ and $i=1,2,3$
describe the spin singlet and the spin triplet, respectively. An explicit
calculation reveals that $r=0,3$ corresponds to the particle-hole channel
(i.e., products ${\bar\psi}\psi$), while $r=1,2$ describes the
particle-particle channel (i.e., products ${\bar\psi}{\bar\psi}$ or
$\psi\psi$). 
From the structure of Eq.\ (\ref{eq:2.8}) one obtains the
following formal symmetry properties of the $Q$ matrices,\cite{us_fermions}
\begin{mathletters}
\label{eqs:2.11}
\begin{eqnarray}
{^0_r Q}_{12}&=&(-)^r\,{^0_r Q}_{21}\quad,\quad (r=0,3)\quad,
\label{eq:2.11a}\\
{^i_r Q}_{12}&=&(-)^{r+1}\,{^i_r Q}_{21}\ ,\ (r=0,3;\ i=1,2,3)\quad,
\label{eq:2.11b}\\
{^0_r Q}_{12}&=&{^0_r Q}_{21}\quad,\quad (r=1,2)\quad,
\label{eq:2.11c}\\
{^i_r Q}_{12}&=&-{^i_r Q}_{21}\quad,\quad (r=1,2;\ i=1,2,3)\quad,
\label{eq:2.11d}\\
{^i_r Q}_{12}^*&=&- {^i_r Q}_{-n_1-1,-n_2-1}^{\alpha_1\alpha_2}\quad.
\label{eq:2.11e}
\end{eqnarray}
\end{mathletters}%
Here the star in Eq.\ (\ref{eq:2.11e}) denotes complex conjugation.

Now by using the delta constraint in Eq.\ (\ref{eq:2.9}) to rewrite all terms 
that are quartic in the fermion field in terms of $Q$, we can achieve
an integrand that is bilinear in $\psi$ and $\bar\psi$. The Grassmannian
integral can then be performed exactly, and we obtain for the effective
action ${\cal A}$
\begin{eqnarray}
{\cal A}[Q,{\wt\Lambda}] &=& {\cal A}_{\rm int}[Q] + {\cal A}_{\rm dis}[Q]
                           + \frac{1}{2}\,\Tr\ln\left(G_0^{-1} - i{\wt\Lambda}
                                       \right)
\nonumber\\
  && + \int d{\bf x}\ \tr\left({\wt\Lambda}({\bf x})\,Q({\bf x})\right)\quad.
\label{eq:2.12}
\end{eqnarray}
Here $\Tr$ denotes a trace over all degrees of freedom, 
including the continuous
position variable, while $\tr$ is a trace over all those discrete indices that
are not explicitly shown. And
\begin{equation}
G_0^{-1} = -\partial_{\tau} + \partial_{\bf x}^2/2m + \mu\quad
\label{eq:2.13}
\end{equation}
is the inverse free electron Green operator, with $\partial_{\tau}$ and
$\partial_{\bf x}$ derivatives with respect to imaginary time and position,
respectively, $m$ is the electron mass, and $\mu$ is the chemical potential.
We can see
from the structure of the $\Tr\ln$-term in Eq.\ (\ref{eq:2.12}) that the
physical meaning of the auxiliary field ${\wt\Lambda}$ is 
that of a self--energy.
The electron-electron interaction 
${\cal A}_{\rm int}$ is conveniently decomposed into four pieces 
that describe the interaction
in the particle-hole and particle-particle spin-singlet and spin-triplet 
channels.\cite{us_fermions} 
For the purposes of the present paper, we need only the
particle-particle spin-singlet channel interaction explicitly to describe 
superconductivity. Similar to the BCS model we ignore the normal Coulomb 
repulsion in the particle-hole channels, and we also ignore the 
possibility of triplet superconductivity.\cite{TRI} Then 
\begin{eqnarray}
{\cal A}_{\rm int}[Q] &=& {\cal A}_{\rm int}^{\,(c)} 
\nonumber\\
    &=&\frac{T\Gamma^{(c)}}{2}\int d{\bf x}\sum_{r=1,2}
                    \sum_{n_1,n_2,m}\sum_{\alpha}
\nonumber\\
&&\times\left[\tr\left((\tau_r\otimes s_0)\,Q_{n_1,-n_1+m}^{\alpha\alpha}
({\bf x})\right)\right]
\nonumber\\
&&\times\left[\tr\left((\tau_r\otimes s_0)\,Q_{-n_2,n_2+m}^{\alpha\alpha}
({\bf x})\right)\right]\quad,
\label{eq:2.14}
\end{eqnarray}
with $\Gamma^{(c)}$ the particle-particle spin-singlet channel 
interaction amplitude, 
with $\Gamma^{(c)} < 0$ leading to superconductivity. 
For the disorder part of the effective action one finds\cite{dis_footnote}
\be
{\cal A}_{\rm dis}[Q] = \frac{1}{\pi N_F\tau_{\rm e}}\int d{\bf x}\
                       \tr \bigl(Q({\bf x})\bigr)^2\quad.
\label{eq:2.15}
\ee

We will focus on the matrix elements $^0_0 Q$ and $^0_1 Q$ in 
disordered superconductivity states. 
From Eqs.\ (\ref{eq:2.8}) and
(\ref{eq:2.10}) we find
\begin{mathletters}
\label{eqs:2.16}
\begin{eqnarray}
{^0_0 Q}_{12}({\bf x})&\cong&\frac{i}{8}\,\left[
        -  \psi_{1\uparrow}({\bf x}){\bar\psi}_{2\uparrow}({\bf x})
   - \psi_{1\downarrow}({\bf x}){\bar\psi}_{2\downarrow}({\bf x})\right.
\nonumber\\
&& + \left.{\bar\psi}_{1\downarrow}({\bf x})\psi_{2\downarrow}({\bf x})
   + {\bar\psi}_{1\uparrow}({\bf x})\psi_{2\uparrow}({\bf x})\right]
                                                     \quad,
\label{eq:2.16a}
\end{eqnarray}
\begin{eqnarray}
{^0_1 Q}_{12}({\bf x})&\cong&\frac{-1}{8}\,\left[
        - \psi_{1\uparrow}({\bf x})\psi_{2\downarrow}({\bf x})
   + \psi_{1\downarrow}({\bf x})\psi_{2\uparrow}({\bf x})\right.
\nonumber\\
&& + \left.{\bar\psi}_{1\downarrow}({\bf x}){\bar\psi}_{2\uparrow}({\bf x})
   - {\bar\psi}_{1\uparrow}({\bf x}){\bar\psi}_{2\downarrow}({\bf x})\right]
                                                     \quad.
\label{eq:2.16b}
\end{eqnarray}
\end{mathletters}%
Note that ${^0_2 Q}_{12}$ has a structure 
similar to ${^0_1 Q}_{12}$. This implies we could 
use ${^0_2 Q}_{12}$ instead of ${^0_1 Q}_{12}$. 
Physically, ${^0_0 Q}_{12}$ is 
related to the single particle density of states, while
${^0_1 Q}_{12}$ is basically the superconducting order parameter.

\section{Saddle--point solutions and Gaussian approximation}
\label{sec:III}

\subsection{The saddle--point method}
\label{subsec:III.A}

We now look for a saddle-point solution of the field theory derived in
the previous section. 
The saddle--point condition is\cite{us_fermions,RMP}
\begin{equation}
\frac{\delta {\cal A}}{\delta Q}\bigg\vert_{Q_{\rm sp},\wt\Lambda _{\rm sp}} 
= \frac{\delta {\cal A}}{\delta\wt\Lambda}\bigg\vert_{Q_{\rm sp},
                      \wt\Lambda _{\rm sp}} = 0\quad.
\label{eq:3.1}
\end{equation}
According to Eqs.\ (\ref{eqs:2.16}), the saddle point values of both 
$Q$ and $\wt\Lambda$ in singlet superconductivity--like 
phases have the structures
\begin{mathletters}
\label{eqs:3.2}
\begin{eqnarray}
{_r^iQ}_{12}({\bf x})\Bigl\vert_{\rm sp}&=&
\delta_{\alpha_1\alpha_2}\,\delta_{i0}\,\left[
\delta_{n_1,-n_2}\,\delta_{r1}\,Q_{n_1}\right.
\nonumber\\
&&+ \left.\delta_{n_1,n_2}
   \,\delta_{r0}\,\Lambda_{n_1}\right]\quad,
\label{eq:3.2a}\\
{_r^i\wt\Lambda}_{12}({\bf x})\Bigl\vert_{\rm sp}&=& 
\delta_{\alpha_1\alpha_2}\,\delta_{i0}\,\left[
\delta_{n_1,-n_2}\,\delta_{r1}\,(i q_{n_1})\right.
\nonumber\\
&&+ \left.\delta_{n_1,n_2}
   \,\delta_{r0}\,(-i\lambda_{n_1})\right]\quad.
\label{eq:3.2b}
\end{eqnarray}
\end{mathletters}%
where we assume $\Lambda_n = - \Lambda_{-n}$, $\lambda_n = 
- \lambda_{-n}$ which is 
equivalent to a redefinition of the chemical 
potential,\cite{RMP} and set $Q_n = Q_{-n}$, 
$q_n = q_{-n}$ which follows from Eqs.\ (\ref{eqs:3.2}) and (\ref{eq:2.11c}).
Substituting this into Eqs.\ (\ref{eq:2.12}) - (\ref{eq:2.15}), and
using the saddle-point condition Eq.\ (\ref{eq:3.1}), 
we obtain the saddle-point
equations
\bml
\label{eqs:3.3}
\bea
\Lambda_n&=&\frac{i}{2V}\sum_{\bf k} {\cal G}_n({\bf k})\quad,
\label{eq:3.3a}\\
Q_n&=&\frac{-i}{2V}\sum_{\bf k} {\cal F}_n({\bf k})\quad,
\label{eq:3.3b}\\
\lambda_n&=&\frac{-2i}{\pi N_{\rm F}\tau_{\rm e}}\,\Lambda_n\quad,
\label{eq:3.3c}\\
q_n&=&\frac{2i}{\pi N_{\rm F}\tau_{\rm e}}\ Q_n - 4i\,\Gamma^{(c)}\,T
                \sum_m Q_m\quad.
\label{eq:3.3d}
\eea
\eml%
Here
\bml
\label{eqs:3.4}
\bea
{\cal G}_n({\bf k})&=&\frac{- (i\omega_n - \lambda_n) - \xi_{\bf k}}
{- (i\omega_n - \lambda_n)^2 + \xi_{\bf k}^2 + q_n^2}\quad,
\label{eq:3.4a}\\
{\cal F}_n({\bf k})&=&\frac{q_n}
{- (i\omega_n - \lambda_n)^2 + \xi_{\bf k}^2 + q_n^2}\quad,
\label{eq:3.4b}
\eea
\eml%
are Green functions 
with $\xi_{\bf k} = {\bf k}^2/2m - \mu$. 

From Eqs.\ (\ref{eqs:3.3}), it is easy to find
\bml
\label{eqs:3.5}
\bea
\lambda_n&=&\frac{1}{\pi N_{\rm F}\tau_{\rm e}}
	\,\frac{1}{V}\sum_{\bf k} {\cal G}_n({\bf k})\quad,
\label{eq:3.5a}\\
q_n&=&\frac{1}{\pi N_{\rm F}\tau_{\rm e}}\,
\frac{1}{V}\sum_{\bf k} {\cal F}_n({\bf k}) 
\nonumber\\
&&- 2\,\Gamma^{(c)}\,T
	\,\frac{1}{V}\sum_{\bf k} \sum_m {\cal F}_m({\bf k})\quad.
\label{eq:3.5b}
\eea
\eml%
We now define a gap function $\Delta$ by\cite{tri_sc0}
\bml
\label{eqs:3.6}
\be
q_n = \bar{q}_n + \Delta \equiv \eta_n\Delta\quad,
\label{eq:3.6a}
\ee
with
\be
\bar{q}_n = \frac{1}{\pi N_{\rm F}\tau_{\rm e}}
\,\frac{1}{V}\sum_{\bf k} {\cal F}_n({\bf k})\quad,
\label{eq:3.6b}
\ee
and it can be shown that
\be
\eta_n\omega_n = i\lambda_n + \omega_n\quad.
\label{eq:3.6c}
\ee
\eml%
We then obtain the gap equation,
\begin{eqnarray}
\Delta &=& - 2\,\Gamma^{(c)}\,T
	\,\frac{1}{V}\sum_{\bf k} \sum_n \frac{\eta_n\Delta}
{(\eta_n\omega_n)^2 + \xi_{\bf k}^2 + (\eta_n\Delta)^2}
\nonumber\\
	 &=& - 2\,\Gamma^{(c)}\,T
	\,\sum_n N(0)\int d{\xi_{\bf k}}\, \frac{\Delta}
{\omega_n^2 + \xi_{\bf k}^2 + \Delta^2}  
\label{eq:3.7}
\end{eqnarray}
with $N(0)=\frac{N_F}{2}$ the density 
of states per spin at the Fermi surface. 
A remarkable aspect of this gap equation is that in this 
approximation the gap $\Delta$ and 
the critical temperature $T_c$ are independent of 
the (nonmagnetic) disorder, and so are all thermodynamic 
properties in superconductivity. 
This result is known as Anderson's theorem.\cite{Anderson}

We next obtain the density of states. From Eq.\ (\ref{eq:2.16a}) it follows, 
\begin{equation}
N(\epsilon_F + \omega) = \frac{4}{\pi}\,{\rm Re}\,\Bigl\langle{^0_0 Q}_{nn}
        ({\bf x})\Bigr\rangle\Bigr\vert_{i\omega_n\rightarrow\omega + i0}
                                                                \quad.
\label{eq:3.8}
\end{equation}
In saddle point approximation, we have for the density of states
\begin{eqnarray}
N(\epsilon_F + \omega) &=& \frac{-2}{\pi}\,\frac{1}{V}\,\sum_{\bf k} {\rm Im}\,
    {\cal G}_n({\bf k},i\omega_n\rightarrow\omega + i0)
\nonumber\\
	&=& N_F\,\frac{\omega}{\sqrt{\omega^2 - \Delta^2}}\quad
		\text{for $\omega > \Delta$}
\nonumber\\
	&=& 0\quad\text{for $\omega < \Delta$}\quad.
\label{eq:3.9}
\end{eqnarray}

For later reference we also define
a matrix saddle-point Green function 
\bml
\label{eqs:3.10}
\be
G_{\rm sp} = \left(G_0^{-1} - i{\wt\Lambda}\right)^{-1}\biggr\vert_{\rm sp}
                \quad,
\label{eq:3.10a}
\ee
whose matrix elements are given by
\bea
(G_{\rm sp})_{nm}({\bf k}) &=& \delta_{nm}\,{\cal G}_n({\bf k})\,
                         (\tau_0\otimes s_0)
\nonumber\\
  && - \,\delta_{n,-m}\,{\cal F}_n ({\bf k})\,(\tau_1\otimes s_0)\quad.
\label{eq:3.10b}
\eea
\eml%
Note that the above results are the standard ones.

\subsection{Gaussian approximation}
\label{subsec:III.B}

We next set up  the calculation of the Gaussian fluctuations 
about the saddle point
discussed above. In the following section these results will 
be used to compute the physical correlation functions in the disordered 
superconducting phase. To this end, we write $Q$ and $\wt\Lambda$ in 
Eqs.\ (\ref{eq:2.12}) - (\ref{eq:2.15}) as,
\begin{mathletters}
\label{eqs:3.11}
\begin{eqnarray}
Q = Q_{\rm sp} + \delta Q\quad,
\label{eq:3.11a}\\
\wt\Lambda = \wt\Lambda_{\rm sp} + \delta\wt\Lambda\quad,
\label{eq:3.11b}
\end{eqnarray}
\end{mathletters}%
and then expand to second or Gaussian order in the fluctuations $\delta Q$ and
$\delta\wt\Lambda$. Denoting the constant saddle point 
contribution to the effective
action by ${\cal A}_{\rm sp}$, and the Gaussian action 
by ${\cal A}_G$, we have, to the Gaussian order, that
\begin{equation}
{\cal A}[Q,{\wt\Lambda}] = 
{\cal A}_{\rm sp} + {\cal A}_G[Q,{\wt\Lambda}]\quad,                
\label{eq:3.12}
\end{equation}
with 
\begin{eqnarray}
{\cal A}_G[Q,\wt\Lambda]&=&{\cal A}_{\rm int}[\delta Q]
                             + {\cal A}_{\rm dis}[\delta Q]
  +\frac{1}{4}\,\Tr\left(G_{\rm sp}\delta{\wt\Lambda}G_{\rm sp}\,
                                               \delta{\wt\Lambda}\right)
\nonumber\\
&&+\int d{\bf x}\ \tr\left(\delta{\wt\Lambda}({\bf x})\,\delta Q({\bf x})
                                                             \right)\quad,
\label{eq:3.13}
\end{eqnarray}

For the quadratic 
part we find
\begin{mathletters}
\label{eqs:3.14}
\begin{eqnarray}
\frac{1}{4}
\Tr\left(G_{\rm sp}\,\delta{\wt\Lambda}\,G_{\rm sp}\,\delta{\wt\Lambda}\right)
  &=& \frac{1}{V}\sum_{\bf k}\sum_{1,2,3,4}\sum_{r,s}\sum_{i,j}
      {^i_r(}\delta{\wt\Lambda})_{12}({\bf k})
\nonumber\\
&&\times\ {^{ij}_{rs}A}_{12,34}({\bf k})\,
          {^j_s(}\delta{\wt\Lambda})_{34}(-{\bf k})\quad.
\nonumber\\
\label{eq:3.14a}
\end{eqnarray}
Here
\begin{eqnarray}
{^{ij}_{rs}A}_{12,34}({\bf k}) &=& \delta_{13}\,\delta_{24}\,
           \varphi^{00}_{12}({\bf k})\,N^{00}_{rs}\,\delta_{ij}\,{_r^i I}_{12}
\nonumber\\
 && + \delta_{13}\,\delta_{2,-4}\,\varphi^{01}_{12}({\bf k})\,N^{01}_{rs}\,
                                         \delta_{ij}\,{_r^i I}_{12}
\nonumber\\
 && + \delta_{1,-3}\,\delta_{24}\,\varphi^{10}_{12}({\bf k})\,N^{10}_{rs}\,
                                         \delta_{ij}\,{_r^i I}_{12}
\nonumber\\
 && + \delta_{1,-3}\,\delta_{2,-4}\,\varphi^{11}_{12}({\bf k})\,N^{11}_{rs}\,
                                         \delta_{ij}\,{_r^i I}_{12} ,
\nonumber\\
&\equiv& {^{ij}_{rs}A}^{(0)}_{12,34}({\bf k})\,{_r^i I}_{12}\quad,
\label{eq:3.14b}
\end{eqnarray}
with $4\times 4$ matrices
\begin{eqnarray}
N^{00}&=&\left(\begin{array}{cc} i\tau_3 & 0 \\
                                    0    & -i\tau_3\\ \end{array}\right)\quad,
\quad
N^{01} = \left(\begin{array}{cc} -i\tau_1 & 0 \\
                                    0     & -i\tau_1\\ \end{array}\right)\quad,
\nonumber\\
N^{10}&=&\left(\begin{array}{cc} -i\tau_1 & 0 \\
                                    0     & i\tau_1\\ \end{array}\right)\quad,
\quad
N^{11} = \left(\begin{array}{cc} -i\tau_3 & 0 \\
                                    0     & -i\tau_3\\ \end{array}\right)\quad,
\nonumber\\
\label{eq:3.14c}
\end{eqnarray}
and
\begin{equation}
{^i_r I}_{12} = 1 + \delta_{12}\,\left[-1 + 
\left({{+\atop +}\atop{+\atop -}}\right)_r\,
  	\left({{+\atop -}\atop{-\atop -}}\right)_i \right]\quad,
\label{eq:3.14d}
\end{equation}
where $\left({{+\atop +}\atop{+\atop -}}\right)_r = \delta_{r0}
+ \delta_{r1} + \delta_{r2} - \delta_{r3}$, etc.
and
\begin{equation}
\varphi^{00}_{nm}({\bf k}) = \frac{1}{V}\sum_{\bf p} {\cal G}_n({\bf p})\,
        {\cal G}_m({\bf p} + {\bf k})\quad,
\label{eq:3.14e}
\end{equation}
\end{mathletters}%
and $\varphi^{01}$, $\varphi^{10}$, and $\varphi^{11}$ defined similarly
with ${\cal G}{\cal G}$ in Eq.\ (\ref{eq:3.14e}) replaced by 
$(-1){\cal G}{\cal F}$, $(-1){\cal F}{\cal G}$, 
and ${\cal F}{\cal F}$, respectively.

In a similar way, the term that couples $\delta{\wt\Lambda}$ and $\delta Q$ can
be written
\begin{mathletters}
\label{eqs:3.15}
\begin{eqnarray}
\Tr\left(\delta{\wt\Lambda}\,\delta Q\right)
  &=& 4\sum_{1,2,3,4}\frac{1}{V}\sum_{\bf k}\sum_{r,i}
      {^i_r(}\delta{\wt\Lambda})_{12}({\bf k})
\nonumber\\
&&\times {^i_r B}_{12}({\bf k})\,
          {^i_r(}\delta Q)_{12}(-{\bf k})\ ,
\label{eq:3.15a}
\end{eqnarray}
where
\begin{equation}
{^i_r B}_{12}({\bf k}) = {_r^i I}_{12}\,
\left({{+\atop -}\atop{-\atop +}}\right)_r\quad.
\label{eq:3.15b}
\end{equation}
\end{mathletters}%

$Q$ and $\wt\Lambda$ can now be decoupled by shifting and scaling the
$\wt\Lambda$ field. If we define a new field ${\bar\Lambda}$ by
\bea
{^i_r(}\delta{\wt\Lambda})_{12}({\bf k}) &=&
	2\,{^{ij}_{rs}(}A^{-1})_{12,34}({\bf k})
\nonumber\\
&&\times\left({^j_s(}\delta{\bar\Lambda})_{34}({\bf k}) 
   - {^j_s(}\delta Q)_{34}({\bf k})\right){^j_s B}_{34} ,
\label{eq:3.16}
\eea
with $A^{-1}$ being the inverse of the matrix $A$ defined in 
Eq.\ (\ref{eq:3.14b}),
then $\bar\Lambda$ and $Q$ decouple. Integrating out the
$\delta{\bar\Lambda}$ fluctuations leads to a  Gaussian action 
completely in terms of $\delta Q$ fluctuations,
\bea
{\cal A}_{\rm G}[Q] &=& 
  - \frac{4}{V}\sum_{\bf k}\sum_{1234}\sum_{rs}\sum_{ij}
  {^i_r(}\delta Q)_{12}({\bf k})\,{^{ij}_{rs}(}A^{-1})_{12,34}({\bf k})
\nonumber\\
&&\times \,{^i_r B}_{12}\,{^j_s B}_{34}\,{^j_s(}\delta Q)_{34}(-{\bf k})
\nonumber\\
&&+ {\cal A}_{\rm int}[\delta Q] + {\cal A}_{\rm dis}[\delta Q]\quad,
\label{eq:3.17}
\eea
It is convenient to rewrite this result as
\bml
\label{eqs:3.18}
\bea
{\cal A}_{\rm G}[Q] &=& \frac{-4}{V}\sum_{\bf k}\sum_{1234}\sum_{rs}\sum_{ij}
   {^i_r(}\delta Q)_{12}({\bf k})\,{^{ij}_{rs}M}_{12,34}({\bf k})\,
\nonumber\\
   &&\times \,{^j_s(}\delta Q)_{34}(-{\bf k})\quad,
\label{eq:3.18a}
\eea
where
\bea
{^{ij}_{rs}M}_{12,34}({\bf k}) &=& {^{ij}_{rs}(}A^{-1})_{12,34}({\bf k})\,
{^i_r B}_{12}{^j_s B}_{34}
\nonumber\\
&&- 2T\Gamma^{(c)}\,\delta_{ij}\delta_{rs}\delta _{1+2,3+4}\,
\left({{0\atop +}\atop{+\atop 0}}\right)_r\,
\left({{+\atop 0}\atop{0\atop 0}}\right)_i 
\nonumber\\
&&-\frac{1}{\pi N_F\tau_{\rm e}}{^i_r B}_{12}\,
\delta_{ij}\delta_{rs}\delta_{13}\delta_{24}\ .
\nonumber\\
\label{eq:3.18b}
\eea
\eml%

\section{Physical correlation functions}
\label{sec:IV}

\subsection{Ultrasonic attenuation by saddle-point approximation}
\label{subsec:IV.A}

We now use the results of the preceding sections to calculate 
transverse ultrasonic attenuation in both clean and disordered 
superconductors. 
As shown in Ref.\ \onlinecite{ua1}, the sound attenuation 
coefficient has the expression
\be
\alpha(\omega) =  \lim_{k\rightarrow 0} 
\frac{\omega}{{\rho_{ion}} c^3_s}\,
   	{\rm Im}\chi({\bf k},i\omega_n\rightarrow\omega + i0)\quad,
\label{eq:4.31}
\ee
where, with $D_x \equiv \partial_{x_1} \partial_{x_2}$,
\bleq
\bea
\chi({\bf k},i\omega_n) &=& 
\frac{1}{m^2_e}\frac{1}{V} \int d{{\bf x}} 
d{{\bf x^{\prime}}} d{{\bf y}} d{{\bf y^{\prime}}}
\exp{(-i {\bf k}\cdot ({\bf x}-{\bf y}))}
\sum_{\sigma_1, \sigma_2} \delta{({\bf x} 
- {\bf x^{\prime}})} \delta{({\bf y} - {\bf y^{\prime}})} D_x D_y 
\nonumber\\	
&&\times\frac{1}{\beta} \sum_{\omega_1, \omega_2} 
\langle \bar{\psi}^{\alpha}_{\omega_1, \sigma_1}({\bf x})
	\psi^{\alpha}_{{\omega_1}-{\omega_n}, \sigma_1}({\bf x^{\prime}}) 
\bar{\psi}^{\alpha}_{\omega_2, \sigma_2}({\bf y})
	\psi^{\alpha}_{{\omega_2}+{\omega_n}, \sigma_2}({\bf y^{\prime}}) 
\rangle\quad.
\label{eq:4.32}
\eea
By introducing a source term of the form 
\be
\delta \tilde{S}^{\,\alpha} = 
\int d{\bf x}\sum_{\omega_n} h(\omega_n, {\bf k}) 
	e^{-i{\bf k}\cdot{\bf x}}
	\,\sum_{\omega, \sigma}
\ \bar{\psi}^{\alpha}_{\omega, \sigma} ({\bf x})\,D_x\,
	\psi^{\alpha}_{\omega+\omega_n, \sigma} ({\bf x})  ,
\label{eq:4.33}
\ee
we can obtain
\be
\chi({\bf k}=0,i\omega_n) = \frac{1}{m^2_e \beta V} \frac{\partial^2 \wt{Z}}
	{\partial{h(\omega_n, {\bf k})} 
	\partial{h(\omega_{-n},{\bf {-k}})}}\biggr\vert_{h=0}
\label{eq:4.34}
\ee
with the third term of the right side of the Eq.\ (\ref{eq:2.12}) becoming 
\bea
{\cal A}_3 &=& \frac{1}{2}\,\Tr\ln\left(G_0^{-1} - i{\wt\Lambda} + D\right)
\nonumber\\
	&=& \frac{1}{2}\,\Tr\ln\left(G_0^{-1} - i({\wt\Lambda}_{sp} 
+ \delta{\wt\Lambda}) + D\right)
\nonumber\\
	&=& \frac{1}{2}\,\Tr\ln\left(1+ D G_{sp} 
- i \delta{\wt\Lambda} G_{sp}\right) 
	+\frac{1}{2}\,\Tr\ln\left(G^{-1}_{sp}\right)
\label{eq:4.35}
\eea
and $D \equiv \sum_{\omega_n} \delta{(\omega_1 - \omega_2 + \omega_n)} 
h \exp{(-i{\bf k}\cdot{\bf x})} D_x$. 
\eleq

In the saddle-point approximation, we 
neglect the  $\delta{\wt\Lambda}$ item and 
have
\be
{\cal A}_3 = \frac{-1}{4}\,\Tr\left(D G_{sp} D G_{sp}\right) + const.
\label{eq:4.36}
\ee
We then obtain the ultrasonic attenuation coefficient, for small frequency,  
\be
\alpha_s (\omega) = \alpha_n \,\frac{2}{1 + \exp{(\beta \Delta)}} 
\label{eq:4.37}
\ee
for both clean and disordered superconductors. 
Here $\alpha_n$ is the attenuation coefficient of the normal metal.\cite{pp} 
In the clean metal it has
\bml
\label{eqs:4.38}
\be
\alpha_{n, clean} = \frac{k_f^4 \omega^2}{30 \pi q {\rho_{ion}} c^3_s}
\label{eq:4.38a}
\ee
with the usual conditions $\omega < q v_f < \Delta$ satisfied. 
In the disordered case 
\be
\alpha_{n, disordered} = 
\frac{2 N(0) k_f^4 \omega^2 \tau}{15 m^2 {\rho_{ion}} c^3_s} ,
\label{eq:4.38b}
\ee
\eml%
where the approximation 
of $\tau_{\rm e} \Delta \ll 1$ is assumed, which is called the dirty 
limit.\cite{DirtyLimit}  
The above result confirms the one of
Levy's which was obtained by Boltzmann's transport equation.\cite{Levy} 
It is noted that no Green function method has been used to obtain this 
result before.

The above method can be used to obtain other physical properties, 
like longitudinal electrical conductivity. 
In that case, higher-order corrections must be included due to the 
gauge invariance problem.\cite{gi} Below we show how to correctly 
obtain the conductivity by using the Gaussian fluctuations about 
the saddle point.

\subsection{Physical correlation functions by Gaussian fluctuations}
\label{sec:IV.B}

\subsubsection{Gaussian propagators}
\label{subsec:IV.B.1}

We now expand our method to calculate the Gaussian 
propagators and then to obtain the number density susceptibility, 
$\chi_n$, the spin density susceptibility, $\chi_s$ and the conductivity.
We find in Appendix \ref{app:A} that the number density susceptibility, 
$\chi_n$, and the spin density susceptibility, $\chi_s$, can be 
expressed in terms
of the $Q$-correlation functions,
\begin{eqnarray}
\chi^{(i)}({\bf k},\omega_n) &=& 
\frac{16T}{V}\sum_{1,2}\sum_{r=0,3}\left\langle
{_r^i(\delta Q)}_{1+n,1}({\bf k})\right.
\nonumber\\
&&\times\left. {_r^i(\delta Q)}_{2+n,2}(-{\bf k})
      \right\rangle\quad , 
\label{eq:4.1}
\end{eqnarray}
with $\chi^{(0)} = \chi_n$ and
$\chi^{(1,2,3)} = \chi_s$. 
Here the Gaussian
propagators in the Eq.\ (\ref{eq:4.1}) are given in 
terms of the inverse of the matrix
$M$ defined in Eq.\ (\ref{eq:3.18b}) by
\begin{eqnarray}
\Bigl\langle{_r^i(\delta Q)}_{12}({\bf k}_1)\,{_s^j(\delta Q)}_{34}
   ({\bf k}_2)\Bigr\rangle_G &=&\frac{V}{8}\,
     \delta_{{\bf k}_1,-{\bf k}_2}
\nonumber\\
 &&\times\,{^{ij}_{rs}M}_{12,34}^{-1}({\bf k}_1)\ ,
\label{eq:4.2}
\end{eqnarray}
where $\left\langle\ldots\right\rangle_G$ denotes an average
with the Gaussian action ${\cal A}_G$. 
We see from Eqs.\ (\ref{eq:4.1}) and (\ref{eq:4.2}) that 
$M^{-1}$ determines the correlation functions within Gaussian approximation.

In the following section we will be interested in the number 
density susceptibility $\chi_n$. Other correlation functions 
can be obtained similarly by 
applying the technique introduced below.
From the expression of $Q$ in terms of the fermion fields, Eq.\ (\ref{eq:2.8}),
it is easy to see that the contributions to Eq.\ (\ref{eq:4.1}) from
$r=0$ and $r=3$ are identical for $\omega_n\neq 0$.
We can therefore write
\begin{equation}
\chi_n({\bf k},\omega_n) = 4T\sum_{1,2}
   {^{00}_{33}M}^{-1}_{1+n,1;2+n,2}({\bf k})\quad,
\label{eq:4.3}
\end{equation}
To find 
$\sum_{1,2}{^{00}_{33}M}^{-1}_{1+n,1;2+n,2}$, we rewrite $M$ as
\begin{eqnarray}
{^{ij}_{rs}M}_{12,34}({\bf k}) &\equiv& {^{ij}_{rs}(}A^{-1})_{12,34}({\bf k})\,
{^i_r B}_{12}{^j_s B}_{34}
\nonumber\\
&& - {^{ij}_{rs}D}_{12,34}
\nonumber\\
&\equiv& {^{ij}_{rs}(}C^{-1})_{12,34}({\bf k})
\nonumber\\
&& - {^{ij}_{rs}D}_{12,34}\quad.
\label{eq:4.4}
\end{eqnarray}
Then we find
\begin{equation}
M^{-1} = \left(C^{-1} - D\right)^{-1} \quad.
\label{eq:4.5}
\end{equation}
It is convenient to write
the inverse of the matrix $M$ as an integral equation,
\begin{equation}
M^{-1} = C + C\,D\,M^{-1}\quad,
\label{eq:4.6}
\end{equation}
with
\begin{equation}
{^{ij}_{rs}C}_{12,34} = {^{ij}_{rs}A}^{(0)}_{12,34}
\left({{+\atop -}\atop{-\atop +}}\right)_r\,{^j_s B}_{34}\quad.
\label{eq:4.7}
\end{equation}
For further simplicity, we set $\Gamma = 2T\Gamma^{(c)}$, 
$\tau^0 = \pi N_F\tau_{\rm e}$ and ${^i_r I}_{12} = 1$ for $\omega_n\neq 0$. 
Expanding Eq.\ (\ref{eq:4.6}) we have 
\bleq
\begin{eqnarray}
{^{00}_{33}M}^{-1}_{12,34} = {^{00}_{33}A}^{(0)}_{12,34} - \Gamma
    \left({{-\varphi^{01}_{12}
   \sum_{78}\delta_{1-2,7+8}\,{^{00}_{23}M}^{-1}_{78,34}}
      \atop{+\varphi^{10}_{12}
               \sum_{78}\delta_{-1+2,7+8}\,{^{00}_{23}M}^{-1}_{78,34}}}\right)
 + \frac{1}{\tau^0}
    \left({{{+\varphi^{00}_{12}\,\,{^{00}_{33}M}^{-1}_{1,2;3,4}}\atop 
            {-\varphi^{01}_{12}\,\,{^{00}_{23}M}^{-1}_{1,-2;3,4}}}
      \atop{{+\varphi^{10}_{12}\,\,{^{00}_{23}M}^{-1}_{-1,2;3,4}}\atop 
            {+\varphi^{11}_{12}\,\,{^{00}_{33}M}^{-1}_{-1,-2;3,4}}}}\right)
\quad, 
\label{eq:4.8}
\end{eqnarray}
where we have used the structures of $B$, Eq.\ (\ref{eq:3.15b}) 
and $A^{(0)}$, Eq.\ (\ref{eq:3.14b}). 
${^{10}_{23}M}^{-1}$ in turn obeys the integral equation
\begin{eqnarray}
{^{00}_{23}M}^{-1}_{12,34} = -{^{00}_{23}A}^{(0)}_{12,34} + \Gamma
    \left({{-\varphi^{00}_{12}
   \sum_{78}\delta_{1+2,7+8}\,{^{00}_{23}M}^{-1}_{78,34}}
      \atop{-\varphi^{11}_{12}
               \sum_{78}\delta_{-1-2,7+8}\,{^{00}_{23}M}^{-1}_{78,34}}}\right)
 - \frac{1}{\tau^0}
    \left({{{-\varphi^{00}_{12}\,\,{^{00}_{23}M}^{-1}_{1,2;3,4}}\atop 
            {-\varphi^{01}_{12}\,\,{^{00}_{33}M}^{-1}_{1,-2;3,4}}}
      \atop{{+\varphi^{10}_{12}\,\,{^{00}_{33}M}^{-1}_{-1,2;3,4}}\atop 
            {-\varphi^{11}_{12}\,\,{^{00}_{23}M}^{-1}_{-1,-2;3,4}}}}\right)
\quad. 
\label{eq:4.9}
\end{eqnarray}
\eleq

Similar results can be obtained for ${^{00}_{33}M}^{-1}_{-1,-2;3,4}$ and 
${^{00}_{23}M}^{-1}_{-1,-2;3,4}$. 

We can now obtain the $\sum_{1,2}{^{00}_{33}M}^{-1}_{1+n,1;2+n,2}$ now. 
Obviously $\sum_{1,2}{^{00}_{23}M}^{-1}_{1+n,-1;2+n,2}$ and 
$\sum_{1,2}{^{00}_{23}M}^{-1}_{-1-n,1;2+n,2}$ need to be determined first. 
We find in Eq.\ (\ref{eq:4.9}) that 
all of the ${^{00}_{23}M}^{-1}$ form a linear 
equation group which can be solved by using Cramer's Rule. 
It is then easy to obtain ${^{00}_{33}M}^{-1}_{1+n,1;2+n,2}$ 
by Eq.\ (\ref{eq:4.8}). Through Eq.\ (\ref{eq:4.3})
the number density susceptibility $\chi_n$ can finally be 
evaluated explicitly. 
Note that 
this technique can then be generalized to obtain all 
elements of $M^{-1}$, which
in turn gives the Gaussian propagators or physical 
correlation functions completely.

\subsubsection{Correlation functions in the clean limit}
\label{subsec:IV.B.2}

In this section we discuss the clean limit, or the non-impurity electron 
gas. Let us perform the clean
limit, $\tau_{\rm e}\rightarrow\infty$. ${\cal A}_{\rm dis}$ then vanishes. 
That also means 
$\lambda_n\rightarrow 0$, $\Lambda_n\rightarrow 0$ and $q_n=\Delta$.

For 
small $\vert {\bf k}\vert$ and $\omega_n$, 
we obtain the number density susceptibility of 
clean superconductor, 
\begin{equation}
\chi_n({\bf k},\omega_n) = - N_F\,\frac{\frac{v^2_f}{3}{\bf k}^2}
   {\omega^2_n + \frac{v^2_f}{3}{\bf k}^2}\quad.
\label{eq:4.15}
\end{equation}

The electrical conductivity $\sigma$ is determined by $\chi_n$ via\cite{Pines}
\begin{equation}
\sigma({\bf k},\omega) = i e^2\frac{\omega}{{\bf k}^2}\,
   \chi_n({\bf k},i\omega_n\rightarrow\omega + i0)\quad.
\label{eq:4.16}
\end{equation}
In particular, the real part of the conductivity as a 
function of real frequencies 
has a delta-function
contribution
\begin{eqnarray}
{\rm Re}\ \sigma(\omega) &=& - \lim_{k\rightarrow 0} 
e^2\frac{\omega}{{\bf k}^2}\,
   	{\rm Im}\chi_n({\bf k},i\omega_n\rightarrow\omega + i0)
\nonumber\\
   &=& \frac{e^2\,N_F\,\pi\,v^2_f}{3}\,
                              \delta(\omega)
\nonumber\\
   &=& \frac{n\,\pi\,e^2}{m}\,\delta(\omega)\quad,
\label{eq:4.17}
\end{eqnarray}
with $n = \frac{k_f^3}{3 \pi^2}$ the particle number density. 
This coincides with the result 
already known.\cite{OLD,Tinkham,Transport} 

Similar procedure can be applied to obtain the spin density susceptibility,
by noting that
\bml
\label{eqs:4.18}
\bea
\chi_{s}({\bf k},\omega_n = 0) &=& \frac{16T}{V}\sum_{1,2}\left\langle
{_3^1(\delta Q)}_{1,1}({\bf k})\,{_3^1(\delta Q)}_{2,2}(-{\bf k})
      \right\rangle
\nonumber\\
&=& 2T \sum_{1,2}{^{11}_{33}M}^{-1}_{1,1;2,2}  
\label{eq:4.18a}
\eea
and
\bea
\sum_{1,2}{^{11}_{33}M}^{-1}_{1+n,1;2+n,2} &=& 
	\sum_{1,2}{^{11}_{33}A}^{(0)}_{1+n,1;2+n,2}
\nonumber\\
&=&  \frac{-N_F}{2T}\,\frac{n_n}{n}
\nonumber\\
&& \,\text{for $\omega_n = 0$, $\vert{\bf k}\vert\rightarrow 0$} ,
\label{eq:4.18b}
\eea
where $n = n_s + n_n$, with $n_s$ the density of 
superconducting electrons, $n_n$ the density of 
normal electrons,\cite{Schrieffer}
\be
n_n = n \int_{-\infty}^{\infty} d{\xi_{\bf p}} 
\frac{\exp{(\frac{\sqrt{{\xi_{\bf p}}^2 + \Delta^2}}{T})}}
{T\,(1 + \exp{(\frac{\sqrt{{\xi_{\bf p}}^2 + \Delta^2}}{T})})^2}  .
\label{eq:4.18c}
\ee
\eml%
The result $\chi_{s}(k\rightarrow 0,\omega_n =0) = 
-N_F\frac{n_n}{n}$ is consistent with Yosida's.\cite{Yosida}

The above result means $\chi_s = 0$ at zero temperature. 
This is because a BCS superconductor 
is a perfect diamagnet at $T = 0$.\cite{Yosida} 
The non-zero part comes from the contribution of normal 
electrons at finite temperature,\cite{White} 
since some Cooper pairs are broken into normal electrons at $T\neq 0$.

\subsubsection{Correlation functions in the disordered case}
\label{subsec:IV.B.3}

Now we turn to the disordered case. The approximation 
of $\tau_{\rm e} \Delta \ll 1$ is again assumed. 
Calculations at $T\rightarrow 0$
show 
that in the limit of long wavelength 
and low frequency,
\begin{equation}
\chi_n({\bf k},\omega_n) = 
- N_F\,\frac{\frac{\pi \Delta \tau_{\rm e} v^2_f}{3}{\bf k}^2}
   {\omega^2_n + \frac{\pi \Delta \tau_{\rm e} v^2_f}{3}{\bf k}^2}\quad,
\label{eq:4.19}
\end{equation}
and the real part of the conductivity as a function of real frequencies 
has also a delta-function
contribution
\begin{eqnarray}
{\rm Re}\ \sigma(\omega\rightarrow 0) &=& - \lim_{k\rightarrow 0} 
e^2\frac{\omega}{{\bf k}^2}\,
   	{\rm Im}\chi_n({\bf k},i\omega_n\rightarrow\omega + i0)
\nonumber\\
   &=& \frac{e^2\,N_F\,\Delta\,\tau_{\rm e}\,\pi^2\,v^2_f}{3}\,
                              \delta(\omega)\quad.
\label{eq:4.20}
\end{eqnarray}
Note that to satisfy the f-sum rule in the disordered case 
the conductivity will not vanish completely at finite frequency. 
Our calculation shows that at $T=0$,
\bea
{\rm Re}\ \sigma(&&\omega > 2\Delta) = \frac{\sigma_n}{\omega} 
\nonumber\\
&&\times\int^{\omega-\Delta}_{\Delta} dE 
\frac{-E(E-\omega)-\Delta^2}
{\sqrt{E^2-\Delta^2} \sqrt{(E-\omega)^2-\Delta^2}} ,
\label{eq:4.51}
\eea
where the vertex corrections resulting from  
the impurity scattering and the interaction have been 
omitted. $\sigma_n$ is the conductivity of normal metal. 
This coincides with the result already known, too.\cite{Transport} 

Again, similar procedure can be applied to obtain the spin 
density susceptibility. We find 
that, at $T = 0$
\be
\chi_{s}(k\rightarrow 0,\omega_n =0) = 0\quad,
\label{eq:4.21}
\ee
That means the spin response in 
the nonmagnetic disordered case is the same as that in the clean limit. 
This is consistent 
with Devereaux and 
Belitz's argument,\cite{PairBrake} which has shown that the nonmagnetic 
disorder 
has no effect on the spin--flip pair breaking 
rate.

\section{Conclusion}
\label{sec:V}

In this paper we presented a method to study the 
transport properties of disordered s--wave superconductor. The crucial idea 
is to first identify the saddle points
of the system by using a symmetry analysis, then to study the fluctuations 
around them to obtain the physical correlation functions. 
The ultrasonic attenuation, 
number density susceptibility, the spin density 
susceptibility and the conductivity have been calculated in the 
clean superconductor, as well as in the disordered superconductor. 
Other properties, like energy correlation function, can be similarly  obtained.
Furthermore, the formalism here can be a powerful tool to study the quantum
phase transitions between normal metal and superconductor.

Finally, we remark, that the techniques used here can be used to 
study gapless s--wave 
superconductors, as well as, for example, disordered d--wave SC 
relevant to the high $T_c$ superconductors.\cite{Thesis}

\acknowledgments

We thank Dietrich Belitz for helpful discussions. This work was 
supported by the NSF under grant numbers DMR-99-75259 and DMR-01-32726. 

\appendix

\section{Correlation functions in terms of $Q$ matrices}
\label{app:A}

The real number density susceptibility has the following 
form\cite{FetterWalecka}
\be
X^R({\bf x}_1 t_1, {\bf x}_2 t_2) = -i\theta(t_1 - t_2)
\langle[{\tilde n}({\bf x}_1 t_1), 
	{\tilde n}({\bf x}_2 t_2)]\rangle 
\label{eq:A.1}
\ee
where
\be
{\tilde n} = n - \langle n \rangle 
\label{eq:A.2}
\ee
with $n$ the number density operator.
It is inconvenient to calculate it directly. Instead, we 
introduce a corresponding 
temperature function that depends on the imaginary--time variables
\be
\chi_n({\bf x}_1 \tau_1, {\bf x}_2 \tau_2) = -\langle T_{\tau}[
{\tilde n}({\bf x}_1 \tau_1) 
{\tilde n}({\bf x}_2 \tau_2)]\rangle
\label{eq:A.3}
\ee
where we have the following relation between Eqs.\ (\ref{eq:A.1}) 
and (\ref{eq:A.3}) with 
the Lehmann representation
\be
X^R({\bf k}, \omega) = \chi_n({\bf k},i\omega_n\rightarrow\omega + i0) .
\label{eq:A.4}
\ee
The time--order indication $T_{\tau}$ of Eq.\ (\ref{eq:A.3}) 
will disappear in the 
functional integral form,\cite{NegeleOrland} which is the case 
in the present paper.

Next we notice that
\begin{mathletters}
\label{eqs:A.5}
\begin{eqnarray}
^{0}_{0}Q_{n_1 n_2}&\cong&\frac{i}{8}
\sum_{\sigma} \left({\bar\psi}_{n_1,\sigma}
   \psi_{n_2,\sigma} + {\bar\psi}_{n_2,\sigma}\psi_{n_1,\sigma}\right) ,
\label{eq:A.5a}\\
^{0}_{3}Q_{n_1 n_2}&\cong&\frac{1}{8}\sum_{\sigma} 
\left({\bar\psi}_{n_1,\sigma}
   \psi_{n_2,\sigma} - {\bar\psi}_{n_2,\sigma}\psi_{n_1,\sigma}\right) .
\label{eq:A.5b}
\end{eqnarray}
\end{mathletters}
By using Eqs.\ (\ref{eq:A.3}) and (\ref{eqs:A.5}) we can then obtain
\begin{eqnarray}
\chi_n({\bf k},\omega_n) &=& 16T\sum_{1,2}\sum_{r=0,3}\left\langle
{_r^0(\delta Q)}_{1+n,1}({\bf k})\right.
\nonumber\\
&&\times\left. {_r^0(\delta Q)}_{2+n,2}(-{\bf k})
      \right\rangle . 
\label{eq:A.6}
\end{eqnarray}

Similar analysis can be applied to find the spin density susceptibility. 
With the spin density
\begin{equation}
{\bf n}_s({\bf k}, \omega_n) = 
\sqrt{\frac{T}{V}} 
\sum_{{\bf p}, \omega}\bigl(\psi({\bf p},\omega),{\bf \sigma}\,
                         \psi({\bf p}+{\bf k},\omega+\omega_n)\bigr)
\label{eq:A.7}
\end{equation}
we can obtain Eq.\ (\ref{eq:4.1}).

\ecols
\end{document}